\documentclass[prb,preprintnumbers,amsmath,amssymb,printfigures,twocolumn]{revtex4}
\usepackage{graphicx}
\usepackage[utf8]{inputenc}
\usepackage{color}
\usepackage[dvips,colorlinks=true,bookmarks,pdftitle={},pdfauthor={},pdfsubject={},pdfkeywords={PDF,LaTeX,hyperlinks,hyperref}]{hyperref}
\usepackage[sort&compress]{natbib}
\usepackage[normalem]{ulem}
\begin{document}
\title{Distinct Correlation between the Vibrational and Thermal Transport Properties of Group \textrm{VA} Monolayer Crystals}
\author{Tu\u{g}bey Kocaba\c{s}}
\affiliation{Department of Advanced Technologies, Graduate School of Sciences, Anadolu University, Eskisehir, TR 26555, Turkey}
\author{Deniz \c{C}ak{\i}r}
\affiliation{Department of Physics and Astrophysics, University of North Dakota, Grand Forks, North Dakota 58202, USA}
\author{O\u{g}uz G\"{u}lseren}
\affiliation{Department of Physics, Bilkent University, Bilkent, Ankara 06800, Turkey}
\author{Feridun Ay}
\affiliation{Department of Electrical and Electronics Engineering, Faculty of Engineering, Anadolu University, Eskisehir, TR 26555, Turkey}
\author{Nihan K Perkg\"{o}z}
\affiliation{Department of Electrical and Electronics Engineering, Faculty of Engineering, Anadolu University, Eskisehir, TR 26555, Turkey}
\author{Cem Sevik}
\affiliation{Department of Mechanical Engineering, Faculty of Engineering, Anadolu University, Eskisehir, TR 26555, Turkey}
\email{csevik@anadolu.edu.tr}
\begin{abstract}
The investigation of thermal transport properties of novel two dimensional materials is crucially important in order to assess their potential to be used in future technological applications, such as thermoelectric power generation. In this respect, lattice thermal transport properties of monolayer structures of the group \textrm{VA} elements (P, As, Sb, Bi, PAs, PSb, PBi, AsSb, AsBi, SbBi, P$_{3}$As$_{1}$, P$_{3}$Sb$_{1}$, P$_{1}$As$_{3}$, As$_{3}$Sb$_{1}$) with black phosphorus like puckered structure were systematically investigated by first principles calculations and an iterative solution of the Phonon Boltzmann transport equation. Phosphorene was found to have the highest lattice thermal conductivity, $\kappa$, due to its low average atomic mass and strong interatomic bonding character. As a matter of course, anisotropic $\kappa$ were obtained for all the considered materials, owing to anisotropy in phonon group velocities and scattering rates (relaxation times) calculated for these structures. However, the determined linear correlation between the anisotropy in $\kappa$ of P, As, and Sb is significant. The results corresponding to the studied compound structures clearly point out that thermal (electronic) conductivity of pristine monolayers might be suppressed (improved) by alloying them with the same group elements. For instance, the room temperature $\kappa$ of PBi along armchair direction was predicted as low as 1.5 Wm$^{-1}$K$^{-1}$, whereas that of P was predicted to be 21 Wm$^{-1}$K$^{-1}$. In spite of the apparent differences in structural and vibrational properties, we peculiarly revealed an intriguing correlation between the $\kappa$ of all the considered materials as $\kappa$=c$_{1}$ + c$_{2}$/$m^{2}$, in particular along zigzag direction. Furthermore, our calculations on compound structures clearly resemble that thermoelectric potential of these materials can be improved by suppressing their thermal properties. The presence of the ultra-low $\kappa$ and high electrical conductivity (especially along the armchair direction) makes this class of monolayers promising candidates for thermoelectric applications.  
\end{abstract}
\maketitle

\section{Introduction}\label{intro}
Triggered by the rediscovery of  black phosphorus (BP), group \textrm{VA} materials, specifically monolayer arsenic (As) and antimony (Sb), have started to receive attention due to their peculiar thermal, optical and electronic properties~\cite{RN813, RN814, RN812,cakir1,cakir2,cakir3}. In fact, this excitement among different researchers is inspired by the fabrication of first phosphorene (P monolayer) based transistor in 2014~\cite{RN823}. The interest not only spreads to the optical and electronic community but also to the research  groups studying thermal properties~\cite{RN814, thermcond6, thermcond1, thermcond5}. From the optical and electronic points of view, phosphorene is appealing because of its high carrier mobility and direct band gap~\cite{RN815}. This band gap is also found to vary from 0.3~eV to 2~eV progressively when its thickness is reduced from its bulk form to single-layer form. On the other hand, the relatively low thermal conductivity and its directional dependency due to the anisotropic lattice structure in particular for phosphorene (P monolayer) and arsenene (As monolayer) is of immense value for thermoelectric applications, providing higher electro-thermal conversion efficiency \cite{RN814, RN818, thermcond6}.  Hence, the materials with such properties present a high potential to start a new line of research and development leading to completely new technologies. 

Group \textrm{VA} elements have the electronic configuration of $s^{2}p^{3}$ in the outermost shell, where this $sp^{3}$ bonding causes a non-planarity, either a puckered or buckled structure~\cite{RN821, thermcond1, C7NR00838D}. Van der Waals forces hold the 2D layers where each atom in a layer is 3-fold coordinated with covalent bonds~\cite{RN822}. The puckered structure of P, As, and Sb have been reported to be thermodynamically stable with high electron/hole mobility \cite{RN826} and low thermal conductivity \cite{RN363} values, which is essential for high-performance innovative thermoelectric research. Therefore, it is very motivating to spend more intense theoretical effort on the  nature of phonon transport in these materials in order to shed light on their lattice thermal transport properties. In addition, low thermal conductivity of the \textrm{VA} column inspires controlling the thermal transport properties by material engineering such as using the alloys of these elements. Hence, in this study, it is also an object of interest regarding the change in thermal conductivity when Group \textrm{VA} elements are considered as compound structures. 

\begin{figure}[!ht]
\includegraphics[width=8 cm]{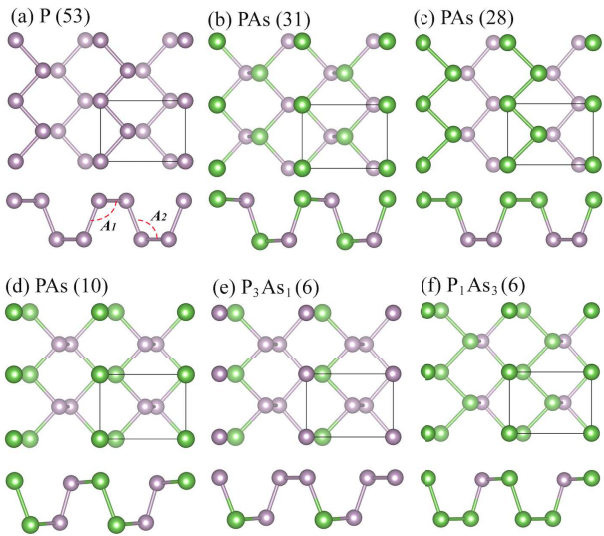}
\caption{\label{structure}Top and side views of the considered two dimensional compounds with different space group symmetries, $Pmna$, $Pmn2_1$, $Pma2$, $P2/m$, and $Pm$.}
\end{figure}
Therefore, we calculate the intrinsic lattice thermal conductivity of 21 stable single layer two dimensional black phosphorane like structures shown in Fig.~\ref{structure}: P, and As with $Pmna$ space group symmetry, Sb and Bi with $Pmn2_1$ space group symmetry, PAs with $Pma2$ space group symmetry,  PAs, PSb, PBi, AsSb, AsBi, SbBi with $Pmn2_1$ and $P2/m$ space group symmetries, and  P$_{1}$As$_{3}$,   P$_{3}$As$_{1}$,  P$_{3}$Sb$_{1}$, and As$_{3}$Sb$_{1}$ with $Pm$ space group symmetry.

\section{Method}\label{metod}
The structural and dynamical properties of all the considered systems were predicted by first-principles calculations based on density functional-theory (DFT) and density-functional perturbation theory (DFPT), as implemented in the Vienna Ab-initio Simulation package (VASP) code~\cite{vasp1, vasp2, PhysRevB.63.155107}. The exchange-correlation interactions were treated using the generalized gradient approximation (GGA) within the Perdew-Burke-Ernzerhof (PBE) formulation\cite{GGA1,Perdew1996}. The single electron wave functions were expanded in plane waves with kinetic energy cutoff of at least 450~eV. For the structure optimizations, the Brillouin-zone integrations were performed using a $\Gamma$-centered regular 21$\times$21$\times$1 $k$-point mesh within the Monkhorst-Pack scheme\cite{Pack1976}. The convergence criterion for electronic and ionic relaxations were set as 10$^{-7}$~eV and 10$^{-3}$~eV/{\AA} , respectively. In order to minimize the periodic interaction along the $z$-direction the vacuum space between the layers was taken at least 15 \r{A}. 

Lattice thermal transport properties were determined by the self-consistent solution of Peierls-Boltzmann transport equation~\cite{BTE} as implemented in ShengBTE code~\cite{ShengBTE1,ShengBTE2}. The zeroth iteration solutions corresponding to Relaxation Time Approximation (RTA) were also predicted in order to clarify the influence of the quasi-momentum-conserving normal and the non-quasi-momentum-conserving Umklapp scattering processes on thermal transport results. The second-order inter-atomic force constants (IFCs) required by the ShengBTE were are obtained by PHONOPY code which extracts the proper force-constant file from the results predicted by DFPT as implemented in VASP. The third-order IFCs were derived from the VASP calculations of the structures generated by considering up to 10 next-nearest-neighbor interactions. Here, 8$\times$8$\times$1 and 5$\times$5$\times$1 supercell structures were used in the DFT calculations performed to predict the second- and third-order IFCs, respectively. In lattice thermal transport calculations at least 40$\times$40$\times$1 well converged $q$-grid and 5.36~{\AA}~\cite{dlength1,dlength2} out of plane lattice constant, $d$ were used for all the considered single layer materials.

\section{Results}\label{results}
\begin{figure}[!ht]
\includegraphics[width=8 cm]{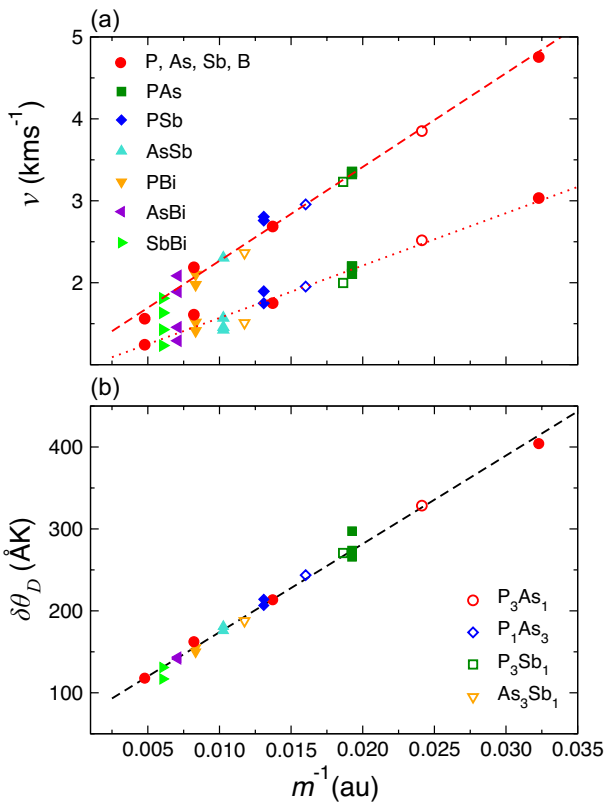}
\caption{\label{figure2} (a) The calculated average acoustic mode group velocities, $v$, along the ZZ and AC directions versus m$^{-1}$. Here, $m$ is the average atomic mass of the unit cell. The red dashed and dotted lines represent the linear fitting values for ZZ and AC directions, respectively. (b) $\delta\theta_D$ versus m$^{-1}$, where $\delta^{2}$, $\theta_D$, and $m$ are the area of the unit cell per atom, Debye temperature, and average atomic mass of each system, respectively. The black dashed line shows the linear fitting values.}
\end{figure}

The primitive unit cell of the considered four pristine (P, As, Sb and Bi), thirteen binary compound (PAs, PSb, PBi, AsSb, AsBi, SbBi) and six one-third compound (P$_{3}$As$_{1}$, P$_{3}$Sb$_{1}$, P$_{1}$As$_{3}$, As$_{3}$Sb$_{1}$) group \textrm{VA} monolayer crystals has a rectangular lattice with a four-atom basis and distinctive space group symmetries, namely $Pmna$ (53), $Pmn2_1$ (31), $Pma2$ (28), $P2/m$ (10), and $Pm$ (6), see Fig.~\ref{structure}. Therefore, except for the three acoustic modes involving the in-phase vibrations of atoms in the unit cell, out-of-phase vibrations of the atoms give rise to nine optical modes for all the structures. The calculated phonon dispersion curves, corresponding to these modes, are depicted in \href{Supplementary Materials.pdf}{ Supplementary Materials (SMs)}, as Figures 2, 3, and 4, respectively, together with the group velocity of each phonon mode as a line color. As seen in these figures, no negative frequencies, indicating the dynamical instability at T=0~K, were found for the most of the considered structures. However, appreciable negative frequencies for the PSb, PBi, AsSb, AsBi, and SbBi structures with $Pma2$ space group symmetry were obtained, meaning that these structures are not dynamically stable. Also, few THz negative frequencies for P$_1$Sb$_3$ and As$_1$Sb$_3$ with $Pm$ space group symmetry were found out and these structures were not considered in thermal transport calculations. The in-plane longitudinal acoustic (LA) and transverse acoustic (TA) modes of all the materials have a linear dependence as $q$ goes zero (i.e. long wave length limit), whereas the out-of-plane acoustic or flexural modes (ZA) have a quadratic dependence in the long wavelength limit as commonly observed in two-dimensional (2D) materials such as graphene. The considered compounds display similar dispersion curves but with distinctive Debye frequencies, resulting in different phonon group velocity ($v_{qi}=\frac{\partial\omega_{qi}}{\partial q}$) values. Fig.~\ref{figure2}(a) shows the calculated average of group velocities of three acoustic modes along zigzag (ZZ) and armchair (AC) directions (corresponding to reduced reciprocal coordinates, $q_{x}$ = 0.05, $q_{y}$ = 0.00 for ZZ, and $q_{x}$ = 0.00, $q_{y}$ = 0.05 for AC). A simple correlation between the average of group velocities of three acoustic modes, $v$ and the average effective mass of elemental configuration of the unit cell, $m$ was established as $v=c_{1} + c_{2}/m$, where $c_{1}$ and $c_{2}$ are constants. Our simple expression is consistent with the fact that the lower the average mass, the higher the group velocity is.

The maximum frequency ($\omega_m$) values at the $\Gamma$-point changes as $\omega_{m,\mathrm{P}}$ $>$ $\omega_{m,\mathrm{As}}$ $>$ $\omega_{m,\mathrm{Sb}}$ $>$ $\omega_{m,\mathrm{Bi}}$ and $\omega_{m,\mathrm{PAs}}$ $>$ $\omega_{m,\mathrm{P_{3}As_{1}}}$ $>$ $\omega_{m,\mathrm{PAs}}$, $\ldots$, $>$ $\omega_{m,\mathrm{SbBi}}$ mainly due to the bond stiffness, different atomic masses for pristine monolayers and average atomic mass of the unit cell for the alloy monolayers. When the bond stiffness becomes weaker (i.e. interatomic force constant becomes smaller) and the mass of atom becomes larger, the phonon frequencies shift to lower energies. Thus, the phonon branches become less disperse with exactly the same order, and $v_{qi}$ of phonon modes markedly decrease with increasing the mass of constituent atom(s). In this context, the Debye temperature, calculated from the contributions of individual acoustic modes (i.e. $\theta_{i}=\hbar\omega_{i, max}/k_{B}$, where $i$=$LA$ and $TA$) is given as~\cite{C5RA19747C}
\begin{equation}
\frac{1}{\theta_D^3} =\frac{1}{2}\left(\frac{1}{\theta_{LA}^3}+\frac{1}{\theta_{TA}^3}\right)\label{1}.
\end{equation}
The calculated Debye temperature ($\theta_D$) values for all the considered monolayer materials are listed in Table \textrm{I} in \href{Supplementary Materials.pdf}{SMs}. While $\theta_D$ is 206 K for phosphorene,  it becomes 50 K for in Bi. 
For 2D crystals, Debye temperature can also be defined as follows\cite{Kittel},
\begin{equation}
\theta_D =\frac{\hbar v_{s}}{k_{B}}\left(\frac{4\pi n}{A}\right)^{1/2}\label{2},
\end{equation}
where $v_{s}$ is the sound velocity, $n$ is the number of atoms in the unit-cell, and $A$ is the area of the unit-cell \cite{conflict}. Based on this formula we calculated the square root of the unit cell area per number of atoms, $\delta^2=A/n$, times Debye temperature (calculated from phonon dispersion via Eq.~(\ref{1})) with respect to inverse average atomic mass of the crystals (i.e. total atomic mass over number of atoms in the unit cell, $m/n$). Quite surprisingly, the predicted Debye temperatures from our calculations is correlated with lattice constants and inverse of the average atomic mass as expected from Eq.~(\ref{2}), see  Fig.~\ref{figure2} (b). Note that, the sound velocity in the plot is considered as inversely proportional with average mass as it is in anharmonic crystals and our results shown in Fig.~\ref{figure2} (a) clearly proves those relations. The obtained perfect $m^{-1}$ correlation in both $v$ and $\theta_D$ indicates that these systems might be quite anharmonic. 

Phonon dispersion curves shown in \href{Supplementary Materials.pdf}{SMs} are not symmetric along the $\Gamma$-X-S and  $\Gamma$-Y-S,  giving rise to  anisotropic phonon group velocities along the zigzag and armchair directions. In fact, all the considered monolayers resemble the orthogonal symmetry resulting in distinctive interatomic bonding structure along the ZZ and AC directions, that induces anisotropic vibrational properties. Interestingly, the dispersion of acoustic modes along the ZZ and AC directions for Sb and Bi structures is not so different as compared to P and As structures, which would give rise to a smaller difference in thermal conductivity along the ZZ and AC directions. The predicted group velocities of LA modes of phosphorene along the ZZ and AC directions (see Figure 2 in \href{Supplementary Materials.pdf}{SMs} are in a good agreement with the previously reported first principles results (8.6 and 4.5~km/s)~\cite{thermcond5} and experimental measurements (9.6 and 4.6~km/s)~\cite{FUJII1982579} for bulk black phosphorus.  The acoustic modes travel at higher speed along the ZZ direction since all monolayers are stiffer along the ZZ direction as compared to the AC direction.
Note that $v_{qi}$ of major heat carrier phonon branches, i.e. acoustic modes, of Sb and Bi both along the ZZ and AC directions near the zone center are about two times smaller than those of black phosphorus as seen in Fig.~\ref{figure2}(a). This notable difference in group velocity values of acoustic branches are fairly associated with the difference in $\kappa$ values of these crystals. 

\begin{figure}[!ht]
\includegraphics[width=8 cm]{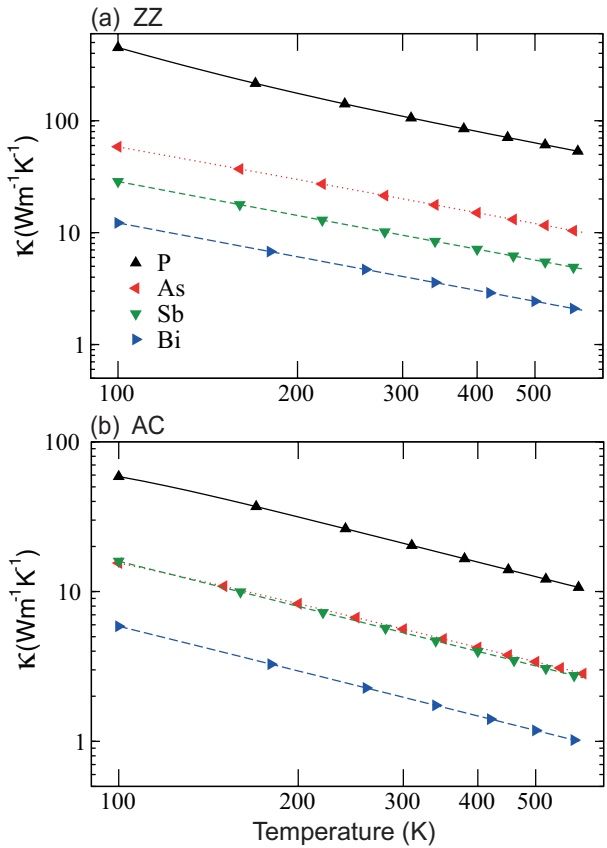}
\caption{\label{figure3} Thermal conductivity of group \textrm{VA} monolayer crystals as a function of temperature along the  (a) zigzag and (b) armchair directions obtained from the self-consistent solution of Peierls-BTE.} 
\end{figure}

On the other hand, the presence of a frequency gap between the low frequency and high frequency optical branches of the most of the considered structure might also be effective on the thermal transport properties of these materials. In fact, in the former studies, the lower thermal conductivity of puckered phosphorus as compared to buckled phosphorus is correlated with the presence of a larger a-o gap in buckled one\cite{thermcond1}. Also, the significant effect of acoustic-optical phonon gaps on the $a+a$ $\leftrightarrow$ $o$ scattering channels was shown for different low dimensional materials~\cite{thermcond1,PhysRevLett.111.025901,PhysRevLett.109.095901,doi:10.1063/1.4893185}. These above-mentioned three phonon processes may also affect the temperature dependence of $\kappa$ since as temperature increases, number of optical phonons increase thereby enhancing three phonon processes involving optical phonons. There is a notable differences in phonon dispersions of compound monolayers with the same concentration but different symmetry, see, for example, Figure 3 in \href{Supplementary Materials.pdf}{SMs}  for the phonon dispersion curves for PSb with symmetries $Pmn2_1$ and $P2/m$. Therefore, we can expect distinctive phonon scattering mechanisms in the compounds with different symmetries. 

Subsequent to phonon dispersion analysis, we first calculated the lattice thermal conductivity of graphene in order to test our methodology and we predicted it as 3290 Wm$^{-1}$K$^{-1}$ which is in quite good agreement with the literature~\cite{Balandin-NM-Rev,sevikTC2,Graphen-Rev1,Graphen-Rev2} (depicted in \href{Supplementary Materials.pdf}{SMs} as Figure 5). Afterwards, we systematically investigated the lattice thermal transport properties of four pristine, thirteen binary and four one-third group \textrm{VA} monolayer structures. The self-consistent solution of Peierls-BTE for the lattice thermal conductivity ($\kappa$) of pristine monolayers as a function of temperature in the range from 100~K to 600~K are depicted with colored symbols in Fig.~\ref{figure3}~(a) and (b).  $\kappa$ decreases along both ZZ and AC directions when moving from P to Bi.  Phosphorene
has the highest $\kappa$ due to its highest $\theta_D$ temperature. We found that all the monolayers exhibit obvious anisotropic thermal transport, which is attributed to the orientation dependent group velocities spotted in Fig.~\ref{figure2} (a) and phonon relaxation times. For instance, the room temperature $\kappa$ along the ZZ direction is 5.21, 3.63, 1.78, and 2.05 times larger than that along the AC direction for P, As, Sb, and Bi, respectively. Interestingly, this ratio, which is nearly the same for Sb and Bi, changes linearly with atomic mass for P, As, and Sb. This behavior is consistent with the recent theoretical works~\cite{thermcond1, thermcond2, thermcond3, thermcond4, thermcond5} performed using first-principles calculations, see Table~\ref{table1}. 

\begin{table}\caption{\label{table1} The calculated room temperature $\kappa$ of the monolayer structures (bold faces), together with the results reported in the literature. $a$, $b$, $c$, $d$, $e$, $f$, $g$, and $h$ corresponds to the results extracted from Ref. \onlinecite{thermcond1}, \onlinecite{C6CP01958G}, \onlinecite{thermcond2}, \onlinecite{thermcond6}, \onlinecite{thermcond3}, \onlinecite{thermcond4}, \onlinecite{thermcond5}, and \onlinecite{PhysRevB.94.165445}, respectively. Here, the values are rescaled with the out of plane lattice constant. 5.36 {\AA} (corresponding to the interlayer distance in bulk black phosphorus) was used in our study to make a reliable comparison of monolayer structures with each other. Out of plane lattice constant values are not available for the results represented with bold italic face.}
\begin{tabular}{llllllll}
\hline\hline
\multicolumn{2}{c}{P}&\multicolumn{2}{c}{As}&\multicolumn{2}{c}{Sb}&\multicolumn{2}{c}{Bi}\\
 $\kappa_{ZZ}$&$\kappa_{AC}$&$\kappa_{ZZ}$&$\kappa_{AC}$&$\kappa_{ZZ}$&$\kappa_{AC}$&&\\\hline
\textbf{109.6}&\textbf{21.0}&\textbf{20.3}&\textbf{5.6}&\textbf{9.6}&\textbf{5.4}&\textbf{4.5}&\textbf{2.7}\\
109.9$^{a}$&23.5$^{a}$&26.8$^{a}$&5.7$^{a}$&9.7$^{a}$&4.7$^{a}$&7.8$^{b}$&6.0$^{b}$\\
\textbf{\textit{30.2}}$^{c}$&\textbf{\textit{13.6}}$^{c}$&\textbf{\textit{30.4}}$^{d}$&\textbf{\textit{7.8}}$^{d}$&&&&\\
\textbf{\textit{48.9}}$^{e}$&\textbf{\textit{27.8}}$^{e}$&&&&&&\\
107.7$^{f}$&35.3$^{f}$&&&&&&\\
81.6$^{g}$&23.8$^{g}$&&&&&&\\
15.3$^{h}$&4.6$^{h}$&&&&&&\\
 \hline\hline
\end{tabular}
\end{table}

Previously, the thermal conductivity of few layers black phosphorus films with a 15~nm thickness was measured as 40 and 20~Wm$^{-1}$K$^{-1}$ and film with a 9.5~nm thickness was measured~\cite{RN814} as 20 and 10~Wm$^{-1}$K$^{-1}$ along ZZ and AC directions, respectively. In addition, the thermal conductivity of a single crystal bulk black phosphorus was measured~\cite{deney2} as 34 and 17~Wm$^{-1}$K$^{-1}$ along the ZZ and AC directions, respectively. Even though we cannot directly compare the $\kappa$ of a monolayer material with that of its few-layer and bulk form, our results for $\kappa$ of phosphorene along both directions are in fair accordance with the previously reported experimental data for bulk and few-layer systems in terms of order of magnitude. Our calculated $\kappa$ of phosphorene is probably higher than that of phosphorene film with a thickness of 9.5 nm due to the strong interlayer interaction in the measured multilayer samples. The overlap of interlayer wavefunctions rather than a pure van der Waals interaction\cite{Tomanek,C5NR06293D} leads to an interlayer interaction much stronger than graphene in this crystal. In addition, the thermal conductivity of phosphorene was calculated by several groups. Interestingly, there is a deviation between the calculated $\kappa$ values as seen in Table~\ref{table1}. This can partially be attributed to different computational methods and computational parameters used in these studies. The calculated values for P, As, and Sb structures in this study agree very well with the values reported in a recent study\cite{thermcond1}. 

\begin{figure}[!ht]
\includegraphics[width=8 cm]{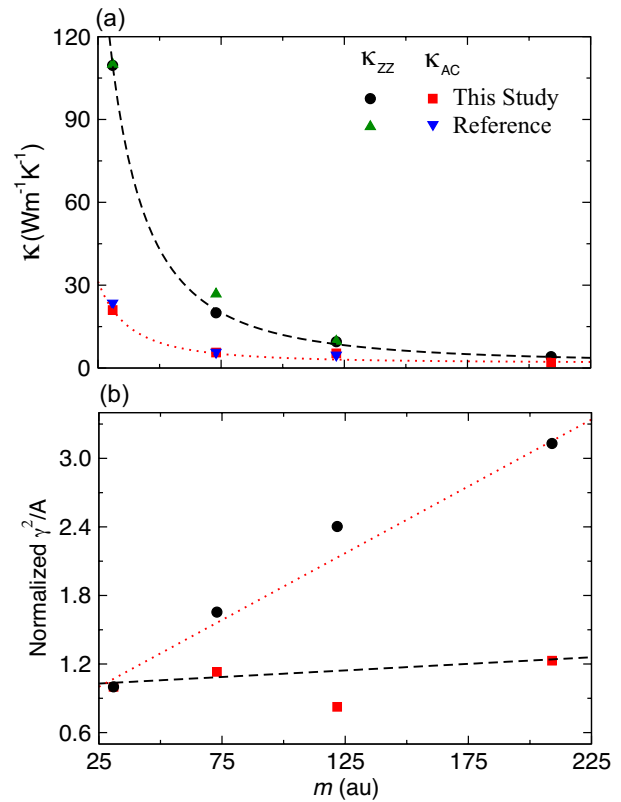}
\caption{\label{figure4} (a) Calculated room temperature $\kappa$ values of the P, As, Sb, and Bi based structures, together with the results reported in Ref.\onlinecite{thermcond1} and (b) the $\gamma^2/A$ as a function of average atomic mass. Here, the black circles and red squares show the values calculated in this study and the green down and blue up triangles show the values reported in the literature~\cite{thermcond1}. The black dashed and red dotted lines in (a) correspond to fitted values (i.e. $\kappa(m)$ = $c_1$+$c_2$/$m^{2}$) along the both directions.}
\end{figure}

Our calculations revealed that there is a distinct correlation between the $\kappa$ of pristine P, As, Sb, and Bi structures and average atomic mass ($m$). We found a simple correlation between $m$ and $\kappa(m)$ such that  $\kappa(m)$ = $c_1$+$c_2$/$m^{2}$, where $c_1$ and $c_2$  are constants, see  Fig.~\ref{figure4}(a). Not surprisingly, the thermal conductivity values predicted by Zheng \textit{et al.} also fits this function very well. In addition to this surprising correlation, we predicted that $\kappa$ of all the pristine monolayers follow nearly the same trend with temperature as seen in Fig.~\ref{figure3}. Therefore, we can certainly claim that the decrease in the $\kappa$ of the pristine structures (i.e. P, As, Sb, and Bi) at temperatures above 100~K follows a $m^{-2}$ behavior except the slight deviation in $\kappa$ of Sb along the AC direction. In fact, the space group symmetry of P and As different from that of Sb and Bi due to the different bond bending angles, $A_1$ and $A_2$ shown in Fig.~\ref{structure} and thus one can naturally expect a distinctive thermal transport properties for these two group. Consequently, the marked correlation revealed as a result of our calculations is quite remarkable.

Previously, Slack suggested that the thermal conductivity ($\kappa_s$) is determined by four factors, including (1) average atomic mass, (2) interatomic bonding, (3) crystal structure and (4) size of anharmonicity\cite{slack}. The first three factors determine harmonic properties.  Three phonon scattering dominated $\kappa_s$ can be given as 
\begin{equation}
\kappa_s =A\frac{m\theta_D^3\delta n^{1/3}}{\gamma^2T}
\end{equation}
where $\delta^2$ is the unit-cell area per atom, $n$ is the number of atoms in the unit cell (which is equal to four for all systems), $A$ is a constant given in terms of average Gr\"{u}neisen parameter of modes~\cite{C5RA19747C}, $\gamma$, and $m$ is the average atomic mass of the unit-cell. We used this equation to predict size of  anharmonicity in these systems. First of all, this simple relation neglects the contribution of the optical modes to the thermal conductivity. This is a reasonable approximation since the contribution of the optical modes to thermal conductivity does not exceeds 10\% for the most of the considered crystals as shown in \href{Supplementary Materials.pdf}{SMs}. Inspired from the above relation, we calculated the $\gamma^2/A$ as,
\begin{equation}
\frac{m\theta_D^3\delta n^{1/3}}{\kappa T},
\end{equation}
at 300 K. Here, we used the $\kappa$ values shown in Fig.~\ref{figure4}.  Then, the results were normalized by the calculated $\kappa$ of P. Within the Slack's approximation, the $\gamma^2/A$ term is directly related with the  anharmonicity of the crystal (the larger the $\gamma$, the larger the anharmonic effect is). Therefore, our calculations clearly show that there is a notable difference between the anharmonic effects along the ZZ and AC directions. While there is a steep increase in the ZZ direction, the values along the AC direction are very close to each other, and the calculated value for Sb is even smaller than that of P. 
\begin{figure*}[!ht]
\includegraphics[width=16 cm]{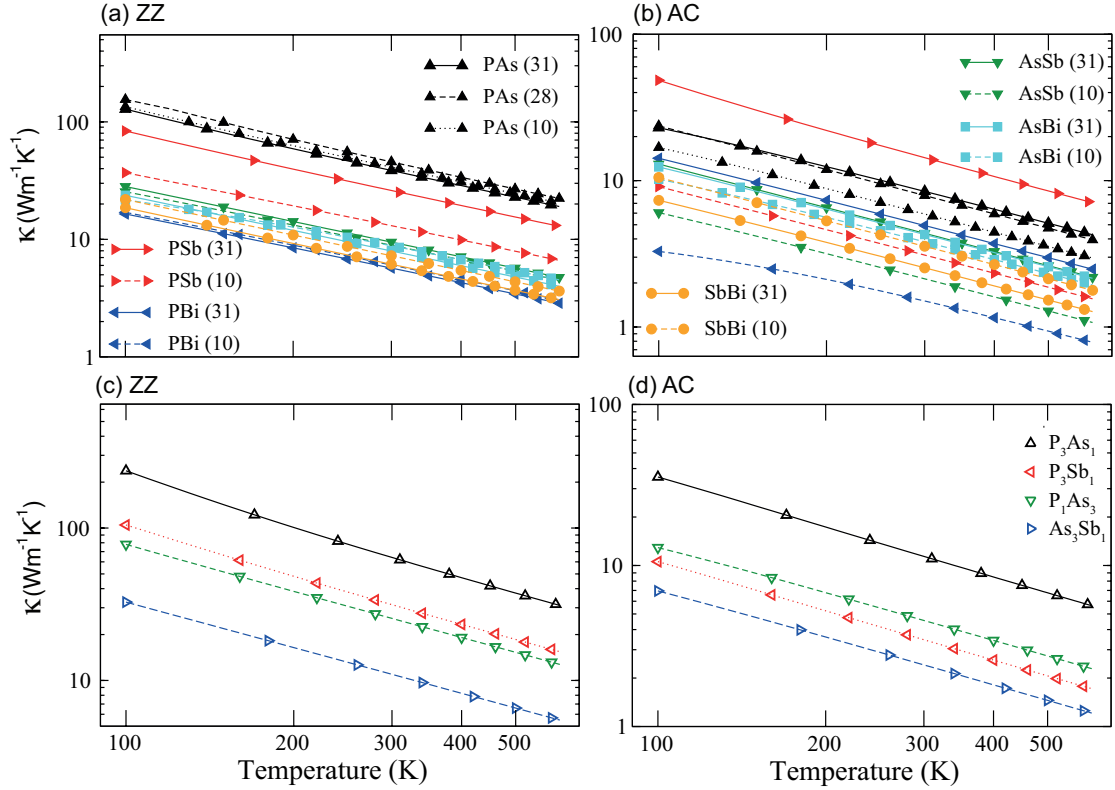}
\caption{\label{figure5} Thermal conductivity of stable group \textrm{VA} compound crystals with all the possible crystal symmetries as a function of temperature along the (a)-(c) zigzag and (b)-(d) armchair directions calculated by the self-consistent solution of Peierls-BTE. Numbers in parenthesis shows the space groups.}
\end{figure*}

Figure~\ref{figure5} shows the calculated lattice thermal transport properties of thirteen binary, (a)-(b), and four one-third, (c)-(d), group \textrm{VA} monolayer structures. Not surprisingly, all the considered compound monolayer crystals exhibit obvious anisotropic thermal transport, which is mainly attributed to the orientation dependent group velocities spotted in Fig.~\ref{figure2}(a). The compound monolayers containing P has larger thermal conductivity as compared with the compounds without P atom. The monolayers with Bi have the lowest thermal conductivities for all temperature values considered here. For instance, $\kappa$ of PBi with $P2/m$ space group was found to be 1.5 Wm$^{-1}$K$^{-1}$. The calculated $\kappa$ of low symmetry compound systems, as low as $\kappa$ of well known 2D thermoelectric materials including single layer SnSe (0.46-0.70  Wm$^{-1}$K$^{-1}$ at 300 K)\cite{thermoelectric}, clearly shows that the thermal conductivity of pristine systems can be suppressed up to order of magnitude by alloying in a controllable manner. Indeed, random elemental distribution in alloy systems results significant reduction in phonon mean free path of long wavelength heat carrier phonons\cite{C0NR00095G,C6NR04651G}. Therefore, it can be reasonably proposed that $\kappa$ of black phosphorus can be tuned (reduced to enhance its thermoelectric potential) by alloying it with Sb, Bi, and also As doping even with very low doping concentrations~\cite{sevikTC1, sevikTC2}, that will lead an opportunity to engineer these materials. This is because of the fact that substitutional As, Sb or Bi doping improves (suppresses) the electronic (thermal) transport by increasing (decreasing) the density of states around the Fermi level (band gap), see Figure 6, 7, and 8 in \href{Supplementary Materials.pdf}{SMs}. In addition, most of the monolayer structures considered here exhibit multiple extrema in their valence and conduction bands. These extrema in the valence and conduction bands are nearly energetically degenerate with the energy difference less than 0.15 eV. The edges of the valence bands yields several hole pockets. These type of electronic features in electronic band structure may give rise to large Seebeck coefficients in these monolayer structures. For the low band gap systems, bipolar conduction may emerge at high temperatures, which lowers  the Seebeck coefficient.  Since the considered monolayer structures have band gap values within range of 0.9-0.09 eV,  we can maximize thermoelectric efficiency by selecting appropriate monolayers.  Combination of the ultra-low $\kappa$ and high electrical conductivity makes these monolayers promising candidates for thermoelectric applications. In our calculations, we did not include spin-orbit coupling (SOC). In fact, 
SOC becomes more effective for monolayer structures with heavy atoms, such as Bi.  It was shown that SOC has a non-negligible
effect on the electronic properties of Bi monolayer\cite{C7RA03662K}. However, we do not expect a significant impact of SOC on phonons and thermal conductivity values. For the calculations of conductivity and Seebeck coefficient, SOC should be taken into account to get accurate results.

\begin{figure}[!ht]
\includegraphics[width=8 cm]{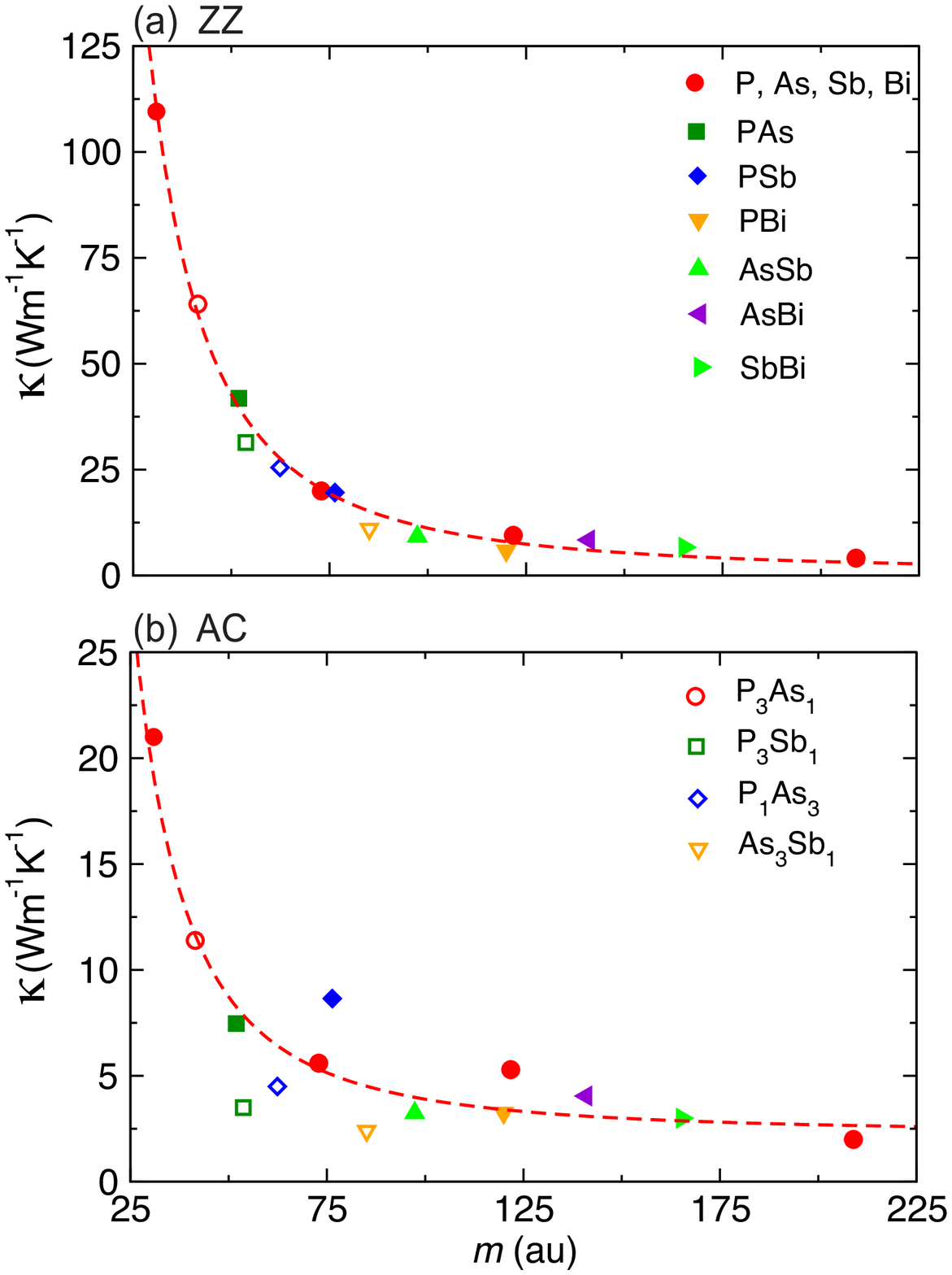}
\caption{\label{contribution} (a) Calculated room temperature $\kappa$ values of the group \textrm{VA} pristine and compound crystals along (a) zigzag and (b) armchair directions. The red dashed lines represent the $c_{1}+c_{2}/m^{2}$ fitting values.}
\end{figure}

The correlation between the $\kappa$ of pristine P, As, Sb, and Bi monolayers and their compound structures with average atomic mass was also searched for all the considered materials. Here, the $\kappa$ for a compound structure was calculated as average over the $\kappa$ values corresponding to all the considered symmetries of an individual concentration. As seen in Fig.~\ref{contribution}(a),  the $\kappa(m)$ = $c_1$+$c_2$/$m^{2}$ curve fits quite well to the calculated $\kappa$ along the ZZ direction of all the considered materials. However, the calculated $\kappa$ along the AC direction of P$_{3}$Sb$_{1}$, PSb, As$_{3}$Sb$_{1}$, and Sb do not follow the same relation as good as along the ZZ direction as seen in Fig.~\ref{contribution}(b). Nevertheless, when these results considered with the identical temperature dependency of $\kappa$ of all the materials represented in Fig.\ref{figure5}, it can be clearly stated that the $\kappa$ of compound group \textrm{VA} monolayer structures can be definitely estimated with $\kappa(m)$ = $c_1$+$c_2$/$m^{2}$ relation, above 100~K, with less than 10\% error, except 36\% and 42\% underestimation for $\kappa$ of Sb and PSb, and 128\% and \%89 underestimation for $\kappa$ of P$_{3}$Sb$_1$ and As$_{3}$Sb$_1$. Despite the apparent differences, such as group velocities and optical phonon gap values, in phonon dispersion of these materials, the revealed simple trend is quite striking. 

\begin{figure}[!ht]
\includegraphics[width=8 cm]{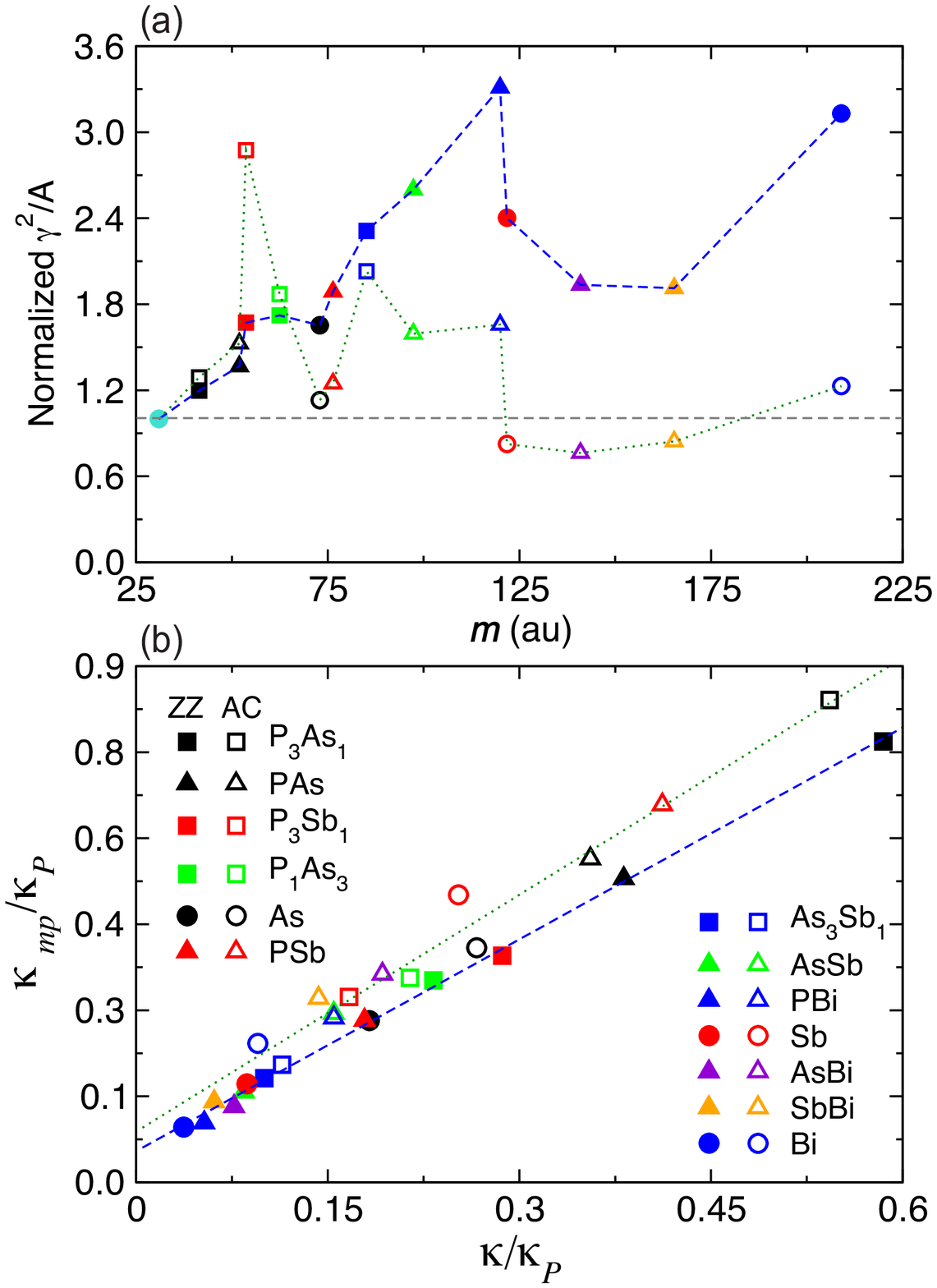}
\caption{\label{normall} (a) The calculated $\gamma^2/A$ values versus average atomic mass, $m$ and (c) $\kappa/\kappa_P$ versus  $\kappa_{mp}/\kappa_P$ for pristine and compound crystals. The blue dashed (solid symbols) and green dotted (open symbols) lines connect the values calculated for zigzag and armchair directions, respectively.}
\end{figure}

\begin{figure}[!ht]
\includegraphics[width=8 cm]{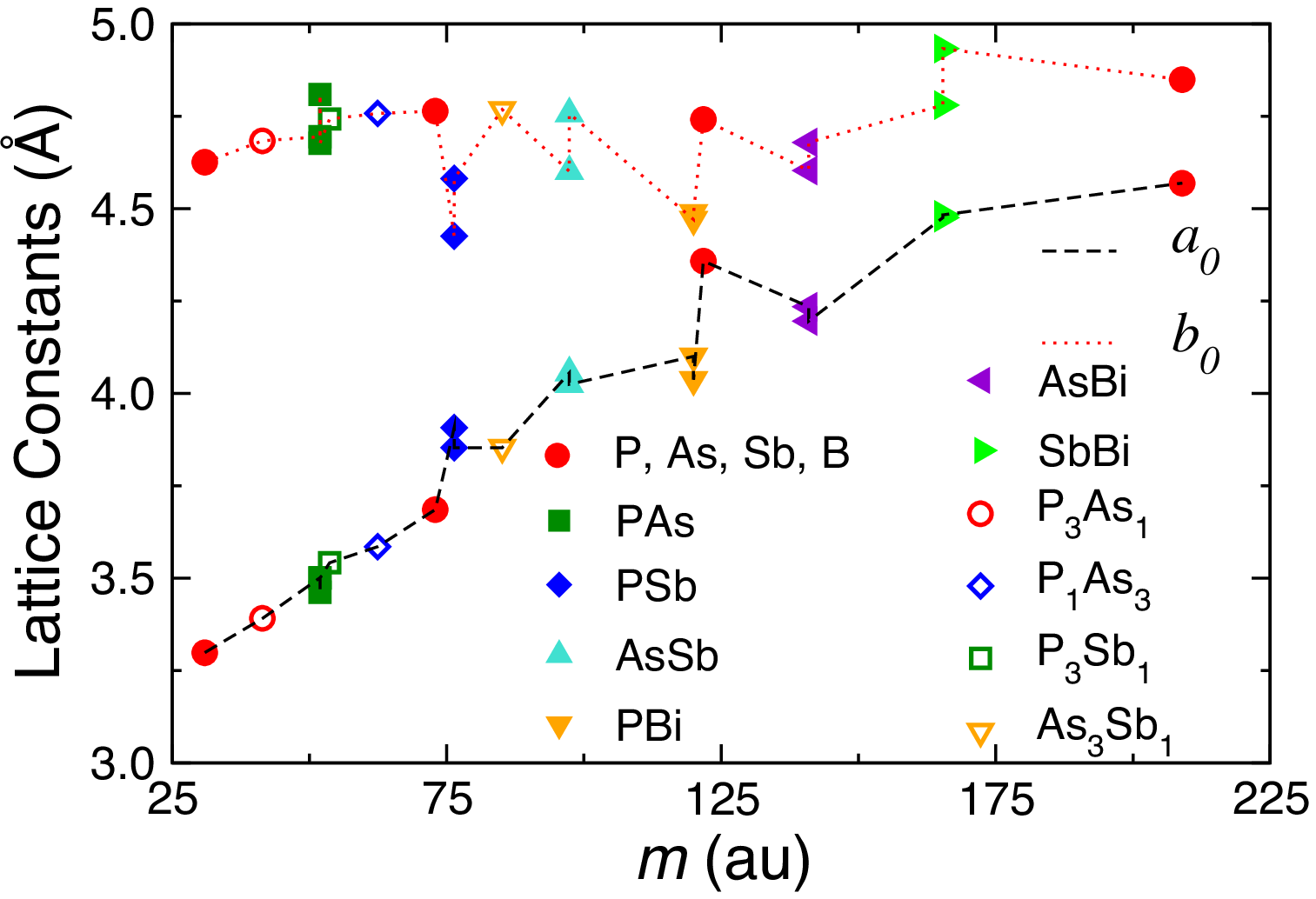}
\caption{\label{lattice} Calculated lattice parameters along ZZ ($a_0$) and AC ($b_0$) directions.}
\end{figure}

The calculated normalized $\gamma^2/A$ values are also calculated for the compound structures. As observed for the pristine monolayers, there is no obvious trend with average atomic mass as seen in Fig.~\ref{normall}(a). However, it can be partially asserted that the normalized conductivity along the ZZ direction increases with increasing average mass, meaning that the effect of anharmonicity becomes more and more prominent. However, the values along the AC direction scatter without an explicit trend. Interestingly, the change in the lattice parameters along the ZZ, $a_0$ and AC, $b_0$ directions, is in line with these results as seen Fig.\ref{lattice}. This partially indicates that the change in some structural parameters, such as $A_1$, $A_2$ and $A_3$ angles depicted in Table \textrm{V} in \href{Supplementary Materials.pdf}{SMs}, due to the change in electronic-ionic bonding nature (i.e P-P is different from Sb-Sb bond), leads to a distinctive direction dependent phonon group velocity and scattering rate values in these monolayer structures. This interpretation is also supported by the mean free path, $l$, dependent thermal conductivities shown in  Figure 9, 10, 11 in \href{Supplementary Materials.pdf}{SMs}. For instance, the mean free path of Sb and  PSb along the AC direction considerably longer than that along the ZZ direction. Conversely, the $l$ of of P, Bi, and PAs ($Pma2$) along  the ZZ direction are rather larger. 

As mentioned above, Slack suggested that the thermal conductivity is determined by four factors\cite{slack}. In this regard, in order to differentiate  the effect of purely interatomic bonding and size of anharmonicity on the thermal conductivity of the crystals, we recalculated the thermal conductivity of all the considered monolayers by equating the mass of all the constituent elements to mass of Phosphor (called as $\kappa_{mp}$). Afterwards, we calculated the ratio of the actual conductivity, $\kappa$, and the one obtained with mass of phosphor, $\kappa_{mp}$ to the thermal conductivity of P, $\kappa_P$, as seen in Figure~\ref{normall}(b). The deviation of $\kappa_{mp}$/$\kappa_P$ from unity depicts the effect of the interatomic bonding and anharmonic interactions on the thermal conductivity. In other words, it shows difference in the strength of interatomic bonds, which results in distinctive vibrational properties and thus anharmonic nature, of all the considered materials. The results for the ZZ direction shows that these two effects dominate more and more the lattice thermal transport  as the average mass of unit cell decreases. However, such correlation for the AC direction is not as strong as the ZZ direction. These results clearly point out the dependence of anharmonicity on the crystal direction. 

The zeroth iteration solutions of Peierls-BTE corresponding to RTA were also calculated. The former is able  to capture simultaneous interaction of all phonon modes, giving rise to collective phonon excitations being responsible for heat transfer.
The ratio of the self-consistent iterative solution to RTA solution for both ZZ and AC directions are depicted in Figure 9, 10, and 11 in \href{Supplementary Materials.pdf}{SMs}, respectively.  The difference between the self-consistent and RTA results is greater than 10\% and independent of temperature when temperature is slightly greater than the $\theta_{D}$ temperature of the monolayers. This  clearly shows that the normal three-phonon processes play an important role in most of the compound crystals investigated in this study.  For the 3D crystals with strong Umklapp scattering, the room temperature $\kappa$ calculated with the iterative solution is typically 10\% greater than RTA solution~\cite{PhysRevB.80.125203} and difference between two methods becomes considerably large for 2D materials. The RTA always severely underestimates thermal conductivity when the normal scattering (being dominant at low T) is strong.  $\kappa$ of all the structures is almost inversely proportional to temperature ($\kappa \sim T^{-1}$), indicating that phonon scattering mechanism is dominated by the Umklapp process.  We found that the dominant scattering mechanisms along ZZ and AC direction is notably different from each other in particular for the structures with P atom. For instance the calculated ratio for the pristine P and As structures along ZZ direction is larger than 1.8, however, the one obtained for the AC direction is smaller than 1.2. The calculated results for pristine Sb explicitly indicate that Umklapp scattering in this structure is considerably stronger than the other pristine ones in particular along the AC direction. This may explain the slight deviation in $\kappa$ of Sb from $\kappa(m)$ = $c_1$+$c_2$/$m^{2}$ relation. The iterative solution adds more than 10\% to $\kappa$ of most of the crystals considered in our study, in the both ZZ and AC directions.  

The ratio of thermal conductivity regarding to the three acoustic and three optical modes to total thermal conductivity, $\kappa_i/\kappa$ (where $i$ = ZA, LA, TA, B$_{1u}$, B$_{1g}$, and B$_{3g}$ and $\kappa$ is total thermal conductivity) was determined for all the materials and the results are represented in Figure 12, 13, and 14 in \href{Supplementary Materials.pdf}{SMs}. The relative contribution of both acoustic and lowest three optical phonons are almost temperature independent for all  materials when the temperature is slightly larger than the $\theta_{D}$ temperature computed using  Eq.~(\ref{1}).  The total contribution of acoustic branches to $\kappa$ is significantly higher than the contribution of optical modes. However, in some P based materials, the lowest three optical modes, B$_{1u}$ and  B$_{1g}$ have a notable contribution as seen in $\kappa_{AC}$ of phosphorene. Also, the contribution of low lying optical mode to conductivity along AC direction is greater than 10\% for most of the P and As based systems (P, As, PAs, AsSb, P$_3$As$_1$,  P$_1$As$_3$, and  As$_3$Sb$_1$). Due to their non-planar structure, monolayers of group \textrm{VA} elements have significantly lower thermal conductivity as compared to graphene, 2000-5000 W/mK~\cite{Balandin-NM-Rev,sevikTC2}. This can be partially attributed to the promoted phonon-phonon scattering of the out-of-plane acoustic (ZA) modes. Due to reflection symmetry perpendicular to in-plane and hexagonal symmetry, three-phonon scattering involving ZA mode (anharmonic scattering of flexural phonons) process is significantly restricted in graphene~\cite{Graphen-Rev1,Graphen-Rev2}. However such a symmetry is broken in group \textrm{VA} monolayers and thus the anharmonic scattering is enhanced in these systems and the contribution of ZA mode not as much as in graphene as clearly seen in mode contribution picture of all the considered monolayers. 

The ZA mode behaves differently along the ZZ and AC directions. While ZA mode provides a significant contribution to total thermal conductivity along the ZZ direction for the most of the monolayer structures, its contribution along the AC direction is usually smaller than 10\% of the total conductivity. This can be explained in terms of distinct structural properties and strength of Umklapp scattering along the ZZ and AC directions. Phosphorene and other group \textrm{VA} monolayers were found to show superior structural flexibility along the AC direction allowing it to have large curvatures. This mechanical flexibility is the origin of  less dispersive ZA, LA  and TA modes along that direction. The quadratic behavior of the ZA mode is more apparent along this direction with a smaller group velocity. In graphene, the contribution of ZA mode to $\kappa$ is 80\% of total conductivity. In general, the contribution of the ZA mode to $\kappa$ is smaller than 50\% in present systems. For all the considered materials,  thermal conductivity of the acoustic modes is more than 70\% of the total thermal conductivity along the ZZ direction. LA and TA modes are always good heat carriers and provide significant contribution to total thermal conductivity along the both ZZ and AC directions.

\section{Conclusion}\label{conc}

The lattice thermal transport properties of group \textrm{VA} monolayers (P, As, Sb, Bi, PAs, PSb, PBi, AsSb, AsBi, SbBi, P$_{3}$As$_{1}$, P$_{3}$Sb$_{1}$, P$_{1}$As$_{3}$, As$_{3}$Sb$_{1}$), found to have no negative frequencies in their phonon dispersions, were systematically investigated by using first principles based calculations. In this manner, the effect of alloying on lattice thermal transport properties of pristine P, As, Sb and Bi were estimated as well. Our calculations revealed that $\kappa$ of all these monolayers can be correlated by their average atomic masses as follows, $\kappa(m) = c_{1}+c_{2}/m^{2}$. In addition, $\kappa$ of all the considered materials are almost inversely proportional to temperature ($\kappa$ $\sim$ T$^{-1}$ ) for the T values, $\geq$ 100 K and therefore the $\kappa(m)$ $\sim$ $m^{-2}$ correlation is clearly applicable for the same T range. Their anisotropic crystal structures give rise to significantly different phonon group velocities and peculiarly scattering rates (phonon relaxation times) along the ZZ and AC directions, thereby anisotropic thermal conductivities.  Using a simple formula developed by Slack\cite{slack}, we revealed that the anharmonic effects, which is becoming more stronger with an increase in the mass of constitute atoms,  play an important role in lowering the thermal conductivity of these monolayers. Our investigations clearly point out that thermal transport properties of phosphorene is tunable via substitutional Sb, As and Bi doping. For instance, the thermal conductivity of P along ZZ direction, 109.6 $Wm^{-1}K^{-1}$, can be suppressed as low as 10 $Wm^{-1}K^{-1}$.  Efficient thermoelectric applications demands materials with ultra low thermal conductivity and good electrical conductivity. In this respect, we can propose PBi as a good candidate due to its ultra low kappa along AC direction, 1.5 Wm$^{-1}$K$^{-1}$.

\section{acknowledgement}
We acknowledge the support from Scientific and Technological Research Council of Turkey (TUBITAK-115F024).  CS acknowledges the support from Anadolu University (BAP-1407F335, -1705F335). and Turkish Academy of Sciences (TUBA-GEBIP). A part of this work was supported by the BAGEP Award of the Science Academy.

\end{document}